\documentclass[aip,apl,twocolumn,superscriptaddress,amsmath,amsfonts,
               floatfix]{revtex4}

\usepackage{epsfig,psfrag}
\usepackage{CJK}

\usepackage{graphicx}
\usepackage{dcolumn}
\usepackage{bm}
\usepackage{amssymb}

\begin{document}
\title{Effect of exciton-phonon coupling on the interlayer excitons in transition metal dichalcogenides double layers}
\author{Zi-Wu Wang}
\affiliation{Tianjin Key Laboratory of Low Dimensional Materials Physics and Preparing Technology,
Department of Physics, Tianjin University, Tianjin 300354, China}
\email{wangziwu@tju.edu.cn}
\author{Xi-Ying Dong}
\affiliation{Tianjin Key Laboratory of Low Dimensional Materials Physics and Preparing Technology,
Department of Physics, Tianjin University, Tianjin 300354, China}
\author{Run-Ze Li}
\affiliation{Tianjin Key Laboratory of Low Dimensional Materials Physics and Preparing Technology,
Department of Physics, Tianjin University, Tianjin 300354, China}
\author{Yao Xiao}
\affiliation{Tianjin Key Laboratory of Low Dimensional Materials Physics and Preparing Technology,
Department of Physics, Tianjin University, Tianjin 300354, China}
\author{Zhi-Qing Li}
\affiliation{Tianjin Key Laboratory of Low Dimensional Materials Physics and Preparing Technology,
Department of Physics, Tianjin University, Tianjin 300354, China}
\begin{abstract}
We investigate the correction of interlayer exciton binding energy in transition metal dichalcogenides double layers arising from the exciton-optical phonon coupling using the method of Lee-Low-Pines unitary transformation. We find that the binding energy varies in several tens of meV, depending on the polarizability of materials and interlayer distance between double layers. Moreover, the correction of binding energy results in the remarkable increasing of the critical temperature for the condensation of dilute excitonic gas basing on the Berezinskii-Kosterlitz-Thouless model. These results not only enrich the knowledge for the modulation of interlayer exciton, but also provide potential insights for the Bose-Einstein condensation and superfluid transport of interlayer exciton in two-dimensional heterostructures.
\end{abstract}

\maketitle

Structure of double layer consisting of different monolayer transition metal dichalcogenides (TMDS), that is called as van der Waals heterostucture\cite{wd1,wd2}, provides an excellent platform to study the interlayer exciton physics\cite{wd3,wd4}, in which an electron and a hole residing in different monolayers, bounded by attractive Coulomb interaction as illustrated in Fig. 1 (a). Due to the spatial charge separation, the recombination lifetimes of interlayer excitons are in the range of nanoseconds, several orders of magnitude longer than intralayer excitons shown in Fig. 1 (b)\cite{wd5,wd6,wd7}. More important is that optical properties of interlayer excitons can be easily modulated by some external ways\cite{wd4,wd6,wd7,wd8,a1}, such as the gate voltage, temperature and power of optical excitation, which provides an ideal candidate for realization of many excitonic devices\cite{wd3,wd9,wd10}, such as excitonic photon storage, excitonic transistor and excitonic light emission dipole. On the other hand, based on excitons are the nature of Bosonic particles, the Bose-Einstein condensation and superfluid transport of the dilute exciton gas in these double layers were proposed extensively in both recent experiments\cite{wd10,wd11} and theories\cite{wd12,wd13,wd14,wd15,wd16}. Of key importance to this species of exciton is its large binding energy, which determines above mentioned potential applications.

Amount of experiments\cite{wd17,wd18,wd19,wd20} and theories\cite{wd14,wd15,wd16,wd19,wd20,wd21} have focused on the binding energies of interlayer excitons in different TMDS double layers in recent years. Wilson et al deduced that the binding energy of interlayer exciton is large than 200 meV in MoSe$_2$-WSe$_2$ heterobilayers by using rational device design and submicrometer angle-resolved photoemission spectroscopy in combination with photoluminescence\cite{wd17}. But Mouri et al found that the exciton binding energy is only 90 meV in MoS$_2$-WSe$_2$ heterobilayers, determining from its thermal dissociation behavior\cite{wd18}. The striking discrepancies between them can be attributed to the differences of the interlayer distance and the effective mass of exciton as theories predicted. Recently, Latini et al developed the quantum electrostatic heterostructure model to calculate the binding energy of interlayer exciton in MoS$_2$-WSe$_2$ bilayers, and found that the consistencies between theory and experiment for the binding energy can be obtained if the properly screening effect of electron-hole interaction is included\cite{wd21}. However, the origin of screening effect as well as other underly determinants of binding energy for interlayer excitons in these TMDS double layers are still ambiguous.
\begin{figure}
\includegraphics[width=3.1in,keepaspectratio]{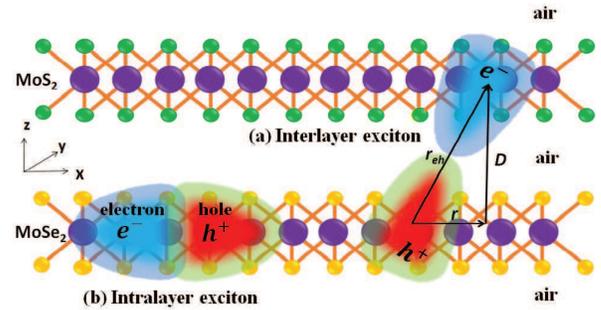}
\caption{\label{compare} Schematic diagrams for the interlayer exciton (a) and intralayer exciton (b) in the MoS$_2$-MoSe$_2$ heterobilayers. $r_{eh}$ is the spatial distance between electron and hole, $D$ is the interlayer distance between different monolayer. The double layer is embedded in the air with the dielectric constant $\varepsilon_d=1$.}
\end{figure}

In the present paper, we mainly study the influence of exciton-longitudinal optical (LO) phonon coupling on the interlayer exciton binding energies in TMDS double layers using the Lee-Low-Pines (LLP) unitary transformation. The corrections of binding energies in different double layers, consisting of the same TMDS (homobilayer) and different TMDS (heterobilayer), are discussed with and without the exciton-LO phonon coupling. We present the dependences of the correction of binding energy on the polarizability of TMDS material and the interlayer distance. In the Berezinskii-Kosterlitz-Thouless (BKT) model, we quantitatively analysis the variation of the critical temperature for the condensation of dilute exciton gas due to the correction of binding energy.

We consider the structure of double layer is composed of two parallel TMDS monolayers with the interlayer distance $D$, in which interlayer exciton is formed that an electron and a hole resides in opposite layers, bounding by the Coulomb interaction schemed in Fig. 1. The total Hamiltonian of interlayer exciton including the intrinsic LO phonons and exciton-LO phonon coupling can be written as
\begin{subequations}
\begin{equation}
H=H_{ex}^{(R)}+H_{ex}^{(r)}+H_{ph}+H_{ex-ph},
\end{equation}
\begin{equation}
H_{ex}^{(R)}=-\frac{\hbar^2}{2M}\nabla_R^2,
\end{equation}
\begin{equation}
H_{ex}^{(r)}=-\frac{\hbar}{2\mu}\nabla_r^2+V(r),
\end{equation}
\begin{equation}
H_{ph}=\sum_{k}\overline{\hbar\omega_{LO}} a_{k}^{\dagger} a_{k},
\end{equation}
\begin{eqnarray}
H_{ex-ph}&=&\sum_{k}\{e^{ik\cdot R}[M_h^je^{-i\beta_1k\cdot r}-M_e^{j'}e^{i\beta_2k\cdot r}] a_{k}\nonumber\\
&&+e^{-ik\cdot R}[M_h^je^{i\beta_1k\cdot r}-M_e^{j'}e^{-i\beta_2k\cdot r}]a_{k}^{\dagger}\}.\nonumber\\
\end{eqnarray}
\end{subequations}
The term $H_{ex}^{(R)}$ denotes the motion of the center-of-mass (CM) an exciton with the mass $M=m_e+m_h$, in which $m_e$ ($m_h$) is the electron (hole) effective mass. The solution of the Schr$\ddot{o}$dinger equation for the motion of CM $H_{ex}^{(R)}\psi(R)=\xi\psi(R)$ is the plane wave $\psi(R)=e^{iQ\cdot R}$ with the quadratic energy spectrum $\xi=\hbar^2Q^2/2M$, where $Q$ is the wave vector of the motion of CM. The second term $H_{ex}^{(r)}$ describes the relative motion of exciton with the reduced mass $\mu=m_e^{-1}+m_h^{-1}$. In general, the relative distance $r$ of electron-hole pair is much smaller than the internal distance $D$, thus the harmonic oscillation approximation $V'(r)=-V_0+\gamma r^2$ for the Coulomb interaction $V(r)=-\kappa e^2/\varepsilon_d\sqrt{D^2+r^2}$ is adopted extensively\cite{wd14,wd15,wd16}, where $V_0=\kappa e^2/\varepsilon_d D$ and $\gamma=\kappa e^2/2\varepsilon_d D^3$ as well as $\varepsilon_d$ is the dielectric constant determined by the dielectric environment and $\kappa=9\times10^9N m^2/C^2$. In this approximation, the eigenfunctions and eigenenergies for the relative motion of interlayer exciton can be solved exactly (see the supplementary material). The obtained binding energy of the ground exciton state is
\begin{equation}
E_b^{(0)}=V_0-\hbar\sqrt{2\gamma/\mu}.
\end{equation}
The third term $H_{ph}$ describes the intrinsic LO phonon and $\overline{\hbar\omega_{LO}}=\sqrt{\hbar\omega_{LO}^j\hbar\omega_{LO}^{j'}}$ is the energy of LO phonon, $j=j'$ and $j\neq j'$ correspond to the homobilayer and heterobilayer, respectively; $a_k$ $(a_{k}^{\dagger})$ is annihilation (creation) a phonon with wave vector $k$. The last term $H_{e-ph}$ denotes the exciton-LO phonon coupling in the Fr$\ddot{o}$hlich mechanism with the coupling element\cite{wd22,wd23,wd24}
\begin{equation}
M_{h(e)}^{j(j')}=\sqrt{\frac{e^2\eta_0 L_m\hbar\omega_{LO}^{j(j')}}{(2A\varepsilon_0 k)}}erfc(\frac{k\sigma}{2}),
\end{equation}
in which $\eta_0$ is the polarization parameter and determined by the polarization properties of the monolayer TMDS, L$_m$ is the monolayer thickness, $erfc$ is the complementary error function and $\sigma$ denotes the effective width of the electronic Bloch states is based on the constrained interaction of LO phonon with charge carriers in monolayer materials, A is the quantization area in the monolayer plane, $\varepsilon_0$ is the permittivity of vacuum. The parameter $\beta_1=m_e/M$ and $\beta_2=m_h/M$ are the fraction of electron and hole, respectively.

Carrying out the LLP unitary transformation\cite{wd25,wd26} for the total Hamiltonian in Eq. (1)
\begin{equation}
\widetilde{H}=U_2^{-1}U_1^{-1}(H_{ex}^{(R)}+H_{ex}^{(r)}+H_{ph}+H_{ex-ph})U_1U_2
\end{equation}
where
\begin{subequations}
\begin{equation}
U_1=\exp\left[i\left(Q-\sum_{k}k a_{k}^{\dagger}a_{k}\right)\cdot R\right],
\end{equation}
\begin{equation}
U_2=\exp\left(\sum_{k}(F_{k} a_{k}^{\dagger} -F_{k}^{\ast}a_{k})\right),
\end{equation}
\end{subequations}
$F_k$ $(F_k^{\ast})$ is the variational function. After the complicate calculation, the transformed Hamiltonian can be rewritten as (see the supplementary material)
\begin{eqnarray}
\widehat{H_{(0)}}&=&\frac{\hbar^2Q^2}{2M}-\frac{\hbar^2}{2\mu}\nabla_r^2+V'(r)+\sum_{k}(\frac{\hbar^2k^2}{2M}+\overline{\hbar\omega_{LO}})a^{\dag}_{k}a_{k}\nonumber\\
&-&\sum_{k}\frac{|M_{h}^j|^2+|M_{e}^{j'}|^2-M_{h}^{j*}M_{e}^{j'}\xi_k(r)}{(\hbar^2k^2/2M)+\overline{\hbar\omega_{LO}}}\nonumber\\
&+&\sum_{k}\frac{\hbar^2k^2}{2\mu}\frac{\beta_1^2|M_{h}^j|^2|+\beta_2^2|M_{e}^{j'}|^2+\beta_1\beta_2M_{h}^{j*}M_{e}^{j'}\xi_k^*(r)}{[(\hbar^2k^2/2M)+\overline{\hbar\omega_{LO}}]^2},\nonumber\\
\end{eqnarray}
with
\begin{equation}
\xi_k(r)=e^{-i(\beta_1+\beta_2)k\cdot r}+e^{i(\beta_1+\beta_2)k\cdot r},\nonumber\\
\end{equation}
in which one phonon term and the interactions between different phonons for CMM and RM are neglected in Eq. (6) due to the tiny correction to the binding energy. In addition, we only discuss the case $Q=0$ in the following, since we are not interested in the CMM.

To obtain the binding energy of excitonic ground state, we choose $|\Phi\rangle=|\phi_{1s}\rangle|0\rangle_{ph}$ as the ground state wave-function of system. $|\phi_{1s}\rangle=1/(\sqrt{2\pi}a)\exp{(-r^2/4a^2)}$ is the exciton ground state function, in which $a=[\hbar/(2\sqrt{2\mu\gamma})]^{1/2}$ is the radius of exciton; $|0\rangle_{ph}$ is the zero phonon state and satisfies the relation of $a_{k}|0\rangle=0$. The corrected exciton binding energy can be obtained via $\widetilde{E_b}=\langle\Phi|\widehat{H_{(0)}}|\Phi\rangle$.
After the mathematical calculations, one can get
\begin{eqnarray}
\widetilde{E_b}&=&E_b^{(0)}-
\sum_{k}\frac{|M_{h}^j|^2+|M_{e}^{j'}|^2-M_{h}^{j*}M_{e}^{j'}\Re(k)}{(\hbar^2k^2/2M)+\overline{\hbar\omega_{LO}}}\nonumber\\
&+&\sum_{k}\frac{\hbar^2k^2}{2\mu}\frac{\beta_1^2|M_{h}^j|^2|+\beta_2^2|M_{e}^{j'}|^2+\beta_1\beta_2M_{h}^{j*}M_{e}^{j'}\Re(k)}{[(\hbar^2k^2/2M)+\overline{\hbar\omega_{LO}}]^2},\nonumber\\
\end{eqnarray}
in which $\Re(k)=\exp{(-a^2k^2/2)}$. Converting the summation of wave vector ($k$) into the integral for Eq. (7), The correction of the exciton binding energy due to the exciton-LO phonons coupling can be analyzed. In the processes of numerical calculations, these average values of $\sigma$=0.55 nm and $L_m$=0.45 nm are adopted for different TMDS materials\cite{wd23,wd24}. The adopted parameters for the LO phonon energies and electron (hole) effective masses are listed in table I. Two distinct types of excitons in TMDS layers, labeled A and B, are formed arising from strong spin-orbit splitting in the valence band. We only illustrate the A type of interlayer exciton in this paper because the similar effect for two types of excitons are assumed.
\begin{table}[htbp]
\caption{\label{compare} Effective masses of electron (hole) and the energies of LO phonons in different TMDS materials are adopted in numerical calculation, which have been taken from Refs. 28 and 29.}
\begin{tabular}{ccccc}
\hline
&MoS$_2$ &  MoSe$_2$  & WS$_2$ &  WSe$_2$\\[1.0ex]\hline
$m_e$($m_0$)       &  0.47  &   0.55   &   0.32  &  0.34  \\
$m_h$($m_0$)&  0.54  &   0.59   &   0.62  &  0.36  \\
$\hbar\omega_{LO}$(meV)         &  47  &   34   &   43  &  30   \\
\hline
\end{tabular}
\end{table}
\begin{figure}
\includegraphics[width=3.3in,keepaspectratio]{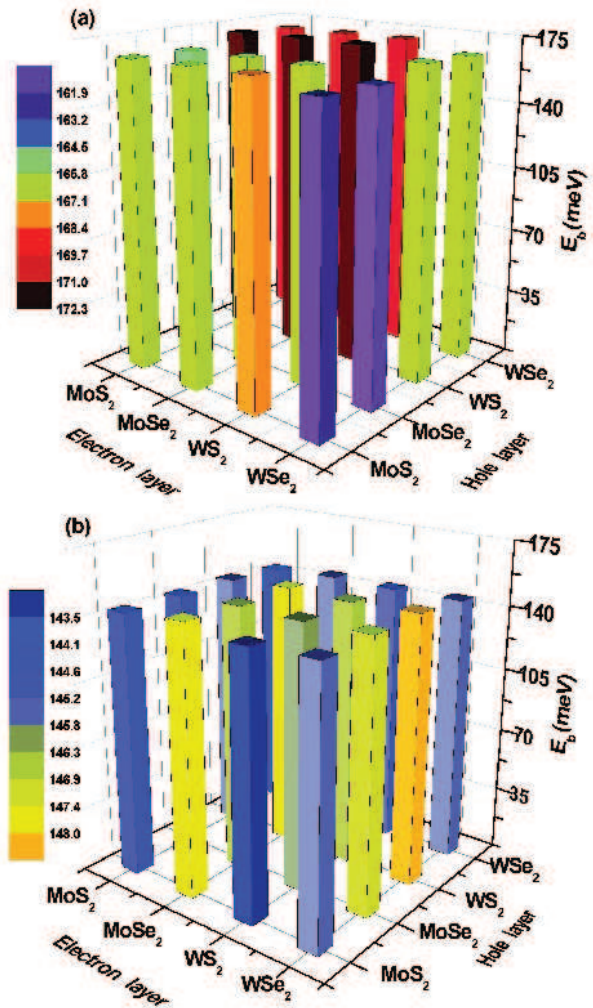}%
\caption{\label{compare} Binding energies of interlayer excitons in different TMDS double layers with (a) and without (b) the exciton-LO phonon coupling.}
\end{figure}
\begin{figure}
\includegraphics[width=3.4in,keepaspectratio]{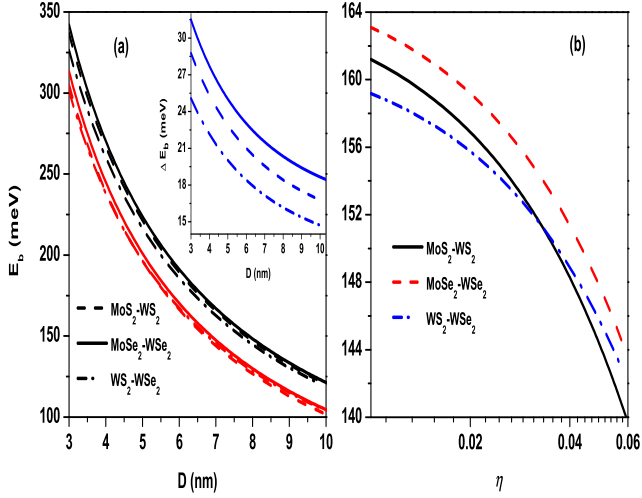}%
\caption{(a) Binding energies as a function of the interlayer distance with (red lines) and without (black lines) exciton-LO phonon coupling in three types of double layers at $\eta$=0.05; The inset presents the dependence of correction of binding energies on the interlayer distance; (b) Binding energies as a function of the polarization parameter in three types of double layers at $D$=7nm.}
\end{figure}
\begin{figure}
\includegraphics[width=3.1in,keepaspectratio]{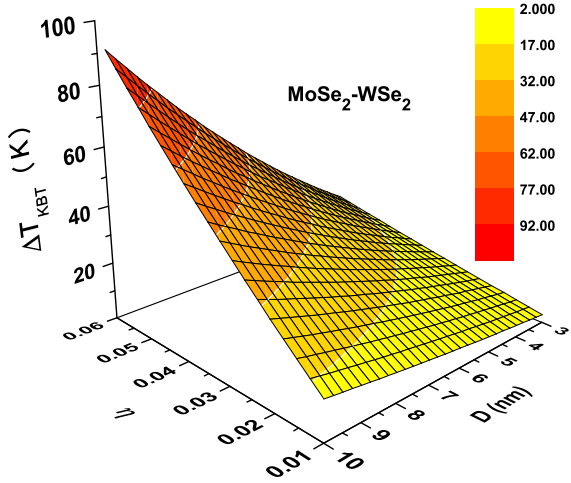}%
\caption{The dependences of the correction of critical temperature on the interlayer distance and polarization parameter in the WSe$_2$-MoSe$_2$ double layer at low exciton density $n=10^{10}cm^{-2}$.}
\end{figure}

According to the distributions of electron and hole in different TMDS monolayer, different interlayer excitons in sixteen types of TMDS double layers, including four homobilayers and twelve heterobilayers, are formed. Comparisons of binding energies for these interlayer excitons are shown in Fig. 2 (a) without the exciton-LO phonon coupling at $D$=7 nm. As can be seen that binding energies are in range of $160\sim173$ meV for these interlayer excitons, which are consistent with the recent prediction\cite{wd29}. The higher binding energies are in MoS$_2$-WS$_2$, MoSe$_2$-WS$_2$ and WS$_2$-WS$_2$ double layers, and the lower ones are in  WSe$_2$-MoS$_2$ and WSe$_2$-MoSe$_2$ double layers. The differences of binding energies between these interlayer excitons can be attributed to the divergencies of effective masses of electron and hole in different TMDS materials. Influences of the exciton-LO phonons coupling on the binding energy are shown in Fig. 2 (b) at $\eta=0.05$. One can see that binding energies decrease in several tens of meV compared to these results presented in Fig. 2 (a). This indicates that the screening effect of Coulomb potential is enhanced by the exciton-LO phonon coupling. In this paper, we have adopted the harmonic approximation for the Coulomb potential describing the electron-hole attraction based on the larger interlayer distances are settled. In fact that anther potential in Keldysh's model has been used extensively\cite{wd14,wd15,wd16}, especially for these double layers with smaller interlayer distance. However, two potentials become closer to each other with increasing the interlayer distance, demonstrating almost no difference as $D>$3 nm.

Fig. 3 (a) presents exciton binding energies as a function of the interlayer distance with and without the exciton-LO phonon coupling in three types of double layers, respectively. One notes that these binding energies decrease in several hundreds of meV as $D$ varies from 3 nm to 10 nm. Therefore, the interlayer distance plays a predominate role to modulate exciton binding energy. Moreover, the corrections of binding energies $\Delta E=\widetilde{E_b}-E_b^{(0)}$ decrease obviously with increasing $D$ shown in the inset of Fig. 3 (a), which means that the strength of exciton-LO phonon coupling can be tuned by the interlayer distance\cite{wd3,wd4}. Fig. 3 (b) presents the corrections of binding energies as a function of the polarization parameter $\eta$ for three different double layers. One find that binding energies reduce markedly with increasing $\eta$. This because of this parameter reflects the strength of exciton-LO phonon coupling directly in Eq. (3). In general, different fixed values were adopted for the coupling strength to fit the experimental measurements\cite{wd30,wd31}. Meanwhile, several recent first-principle calculations have also found that the strength of electron (hole)-LO phonon coupling in Fr$\ddot{o}$hlich mechanism varies in a large scale for different monolayer TMDS materials, depending on the dispersion of LO phonon modes\cite{wd32}. That is the reason we set $\eta$ as a variational parameter to discuss the correction of binding energy.

The Bose-Einstein condensation and superfluidity
of the dilute interlayer excitons gas in these TMDS double layers have aroused intense interesting due to the unusual strong binding energy. According to the BKT model, the critical temperature of interlayer exciton condensation at low densities is given by\cite{wd11,wd12}
\begin{equation}
T_{BKT}=\frac{2.6n}{k_B}\frac{\mu}{M}\frac{e^2}{2\varepsilon_0}a_B^*,
\end{equation}
where $n$ is the exciton density; $k_B$ is the Bolzman constant and $a_B^*$ is the effective exciton Bohr radius, which is related to the exciton binding energy via $a_B^*=e^2/(2\varepsilon_0E_b)$. Then, the correction of the critical temperature in Eq. (8) arising from the exciton-phonon coupling can be given by
\begin{equation}
\Delta T_{BKT}=\frac{2.6n}{k_B}\frac{\mu}{M}(\frac{e^2}{2\varepsilon_0})^2(\frac{1}{\widetilde{E_b}}-\frac{1}{E_b^0}).
\end{equation}
According to Eqs. (2), (7) and (9), the correction of T$_{KBT}$ as functions of the polarization parameter and interlayer distance for WSe$_2$-MoSe$_2$ double layer are shown in Fig. (4) at low exciton density $n=10^{10}cm^{-2}$. It can be seen that: (i) $\Delta$T$_{KBT}$ increases obviously with increasing $D$, which because of the larger $D$ leads to the increase of the exciton effective radius, and consequently the critical value of $na_B^{*2}$ for the dilute exciton condensation; (ii) $\Delta$T$_{KBT}$ increases with increasing $\eta$, which attributes to the increasing of exciton-LO phonon coupling enhances the screening effect, implying the exciton Bohr radius is enlarged. In addition, interlayer excitons in these TMDS double layers embedding with insulator, for example
hexagonal boron nitride (hBN), have been investigated extensively\cite{wd3,wd6,wd8,wd14,wd15,wd16}. In these structures, beside the influence of LO phonon, the excitons coupling with the interface optical phonons induced by the embedding insulator have potential influence on the properties of excitonic states, and need to be explored further.

In summary, we theoretically investigate the influence of exciton-LO phonon coupling on the binding energies of interlayer excitons in different TMDS double layers. We find that exciton binding energies  decrease in some tens of meV due to this kind of many-body interaction. The corrections of binding energies result in the obviously increasing of the critical temperature of the dilute exciton gas condensation. These results not only are very important for the correction of exciton binding energies in TMDS double layers, but also will trigger new experimental studies on van der Waals heterostructures.

This work was supported by National Natural Science Foundation of China (NO 11674241)


\begin{thebibliography}{61}
\expandafter\if
x\csname natexlab\endcsname\relax\def\natexlab#1{#1}\fi
\expandafter\ifx\csname bibnamefont\endcsname\relax
  \def\bibnamefont#1{#1}\fi
\expandafter\ifx\csname bibfnamefont\endcsname\relax
  \def\bibfnamefont#1{#1}\fi
\expandafter\ifx\csname citenamefont\endcsname\relax
  \def\citenamefont#1{#1}\fi
\expandafter\ifx\csname url\endcsname\relax
  \def\url#1{\texttt{#1}}\fi
\expandafter\ifx\csname urlprefix\endcsname\relax\def\urlprefix{URL }\fi
\providecommand{\bibinfo}[2]{#2}
\providecommand{\eprint}[2][]{\url{#2}}

\bibitem[{\citenamefont{{A. K. Geim}}(2013)}]{wd1}
\bibinfo{author}{\bibnamefont{{A. K. Geim and I. V. Grigorieva}}}, \bibinfo{journal}{Nature} \textbf{\bibinfo{volume}{499}}, \bibinfo{pages}{419}
(\bibinfo{year}{2013}).

\bibitem[{\citenamefont{{K. S. Novoselov}}(2016)}]{wd2}
\bibinfo{author}{\bibnamefont{{K. S. Novoselov, A. Mishchenko, A. Carvalho, A. H. Castro Neto}}}, \bibinfo{journal}{Science} \textbf{\bibinfo{volume}{353}}, \bibinfo{pages}{6298}
(\bibinfo{year}{2016}).

\bibitem[{\citenamefont{{H. Fang}}(2014)}]{wd3}
\bibinfo{author}{\bibnamefont{{H. Fang, C. Battaglia, C. Carraro, S. Nemsak, B. Ozdol, J. S. Kang, H. A. Bechtel, S. B. Desai, F. Kronast, A. A. Unal, G. Conti, C. Conlon, G. K. Palsson, M. C. Martin, A. M. Minor, C. S. Fadley, E. Yablonovitch, R. Maboudian, and A. Javey}}}, \bibinfo{journal}{Proc. Natl. Acad. Sci.} \textbf{\bibinfo{volume}{111}}, \bibinfo{pages}{6198}
(\bibinfo{year}{2014}).

\bibitem[{\citenamefont{{P. Rivera}}(2016)}]{wd4}
\bibinfo{author}{\bibnamefont{{P. Rivera, K. L. Seyler, H. Yu, J. R. Schaibley, J. Yan, D. G. Mandrus, W. Yao, and X. Xu}}}, \bibinfo{journal}{Science} \textbf{\bibinfo{volume}{351}}, \bibinfo{pages}{688}
(\bibinfo{year}{2016}).

\bibitem[{\citenamefont{{P. Rivera}}(2015)}]{wd5}
\bibinfo{author}{\bibnamefont{{P. Rivera, J. R. Schaibley, A. M. Jones, J. S. Ross, S. Wu, G. Aivazian, P. Klement, K. Seyler, G. Clark, N. J. Ghimire, J. Yan, D. G. Mandrus, W. Yao, and X. Xu}}}, \bibinfo{journal}{Nat. Commun.} \textbf{\bibinfo{volume}{6}}, \bibinfo{pages}{6242}
(\bibinfo{year}{2015}).

\bibitem[{\citenamefont{{J. S. Ross}}(2017)}]{wd6}
\bibinfo{author}{\bibnamefont{{J. S. Ross, P. Rivera, J. Schaibley, E. L. Wong, H. Yu, T. Taniguchi, K. Watanabe, J. Yan, D. Mandrus, D. Cobden, W. Yao, and X. Xu}}}, \bibinfo{journal}{Nano Lett.} \textbf{\bibinfo{volume}{17}}, \bibinfo{pages}{638}
(\bibinfo{year}{2017}).

\bibitem[{\citenamefont{{B. Miller}}(2017)}]{wd7}
\bibinfo{author}{\bibnamefont{{B. Miller, A. Steinhoff, B. Pano, J. Klein, F. Jahnke, A. Holleitner, and U. Wurstbauer}}}, \bibinfo{journal}{Nano Lett.} \textbf{\bibinfo{volume}{17}}, \bibinfo{pages}{5229} (\bibinfo{year}{2017}).

\bibitem[{\citenamefont{{E. V. Calman}}(2014)}]{wd8}
\bibinfo{author}{\bibnamefont{{E. V. Calman, C. J. Dorow, M. M. Fogler, L. V. Butov, S. Hu, A. Mishchenko,
and A. K. Geim}}}, \bibinfo{journal}{Appl. Phys. Lett.} \textbf{\bibinfo{volume}{108}}, \bibinfo{pages}{101901}
(\bibinfo{year}{2016}).

\bibitem[{\citenamefont{{A. Ciarrocchi}}(2018)}]{a1}
\bibinfo{author}{\bibnamefont{{A. Ciarrocchi, D. Unuchek, A. Avsar, K. Watanabe, T. Taniguchi, A. Kis}}}, \bibinfo{journal}{arXiv:1803.06405}
(\bibinfo{year}{2018}).

\bibitem[{\citenamefont{{F. Withers}}(2015)}]{wd9}
\bibinfo{author}{\bibnamefont{{F. Withers, O. D. Pozo-Zamudio, A. Mishchenko, A. P. Rooney, A. Gholinia, K. Watanabe,
T. Taniguchi, S. J. Haigh, A. K. Geim, A. I. Tartakovskii, and K. S. Novoselov}}}, \bibinfo{journal}{Nat. Mater.} \textbf{\bibinfo{volume}{14}}, \bibinfo{pages}{301}
(\bibinfo{year}{2015}).

\bibitem[{\citenamefont{{L. V. Butov}}(2017)}]{wd10}
\bibinfo{author}{\bibnamefont{{L. V. Butov}}}, \bibinfo{journal}{ Superlattices Microstruct. } \textbf{\bibinfo{volume}{108}}, \bibinfo{pages}{22}
(\bibinfo{year}{2017}).

\bibitem[{\citenamefont{{M. M. Fogler}}(2014)}]{wd11}
\bibinfo{author}{\bibnamefont{{M. M. Fogler, L. V. Butov, and K. S. Novoselov}}}, \bibinfo{journal}{Nat. Commun.} \textbf{\bibinfo{volume}{5}}, \bibinfo{pages}{4555}
(\bibinfo{year}{2014}).

\bibitem[{\citenamefont{{F. C. Wu}}(2015)}]{wd12}
\bibinfo{author}{\bibnamefont{{F. C. Wu, F. Xue, and A. H. MacDonald}}}, \bibinfo{journal}{Phys. Rev. B} \textbf{\bibinfo{volume}{92}}, \bibinfo{pages}{165121} (\bibinfo{year}{2015}).

\bibitem[{\citenamefont{{B. Skinner}}(2016)}]{wd13}
\bibinfo{author}{\bibnamefont{{B. Skinner}}}, \bibinfo{journal}{Phys. Rev. B} \textbf{\bibinfo{volume}{93}}, \bibinfo{pages}{235110} (\bibinfo{year}{2016}).

\bibitem[{\citenamefont{{O. L. Berman}}(2016)}]{wd14}
\bibinfo{author}{\bibnamefont{{O. L. Berman and R. Y. Kezerashvili}}}, \bibinfo{journal}{Phys. Rev. B} \textbf{\bibinfo{volume}{93}}, \bibinfo{pages}{245410}
(\bibinfo{year}{2016}).

\bibitem[{\citenamefont{{O. L. Berman}}(2017)}]{wd15}
\bibinfo{author}{\bibnamefont{{O. L. Berman, G. Gumbs, and R. Y. Kezerashvili}}}, \bibinfo{journal}{Phys. Rev. B} \textbf{\bibinfo{volume}{96}}, \bibinfo{pages}{014505}
(\bibinfo{year}{2017}).

\bibitem[{\citenamefont{{O. L. Berman}}(2017)}]{wd16}
\bibinfo{author}{\bibnamefont{{O. L. Berman and R. Y. Kezerashvili}}}, \bibinfo{journal}{Phys. Rev. B} \textbf{\bibinfo{volume}{96}}, \bibinfo{pages}{094502}
(\bibinfo{year}{2017}).

\bibitem[{\citenamefont{{N. R. Wilson}}(2017)}]{wd17}
\bibinfo{author}{\bibnamefont{{N. R. Wilson, P. V. Nguyen, K. Seyler, P. Rivera, A. J. Marsden, Z. P. L. Laker, G. C. Constantinescu, V. Kandyba, A. Barinov, N. D. M. Hine, X. Xu, and D. H. Cobden}}}, \bibinfo{journal}{Sci. Adv} \textbf{\bibinfo{volume}{3}}, \bibinfo{pages}{e1601832}
(\bibinfo{year}{2017}).

\bibitem[{\citenamefont{{S. Mouri}}(2017)}]{wd18}
\bibinfo{author}{\bibnamefont{{S. Mouri, W. Zhang, D. Kozawa, Y. Miyauchi, G. Edac, and K. Matsuda}}}, \bibinfo{journal}{Nanoscale} \textbf{\bibinfo{volume}{9}}, \bibinfo{pages}{6674}
(\bibinfo{year}{2017}).

\bibitem[{\citenamefont{{D. L. Duong}}(2017)}]{wd19}
\bibinfo{author}{\bibnamefont{{D. L. Duong, S. J. Yun, and Y. H. Lee}}}, \bibinfo{journal}{ACS Nano} \textbf{\bibinfo{volume}{11}}, \bibinfo{pages}{11803}
(\bibinfo{year}{2017}).

\bibitem[{\citenamefont{{T. C. Berkelbach}}(2018)}]{wd20}
\bibinfo{author}{\bibnamefont{{T. C. Berkelbach and D. R. Reichman}}}, \bibinfo{journal}{Annu. Rev. Condens. Matter Phys.} \textbf{\bibinfo{volume}{9}}, \bibinfo{pages}{379}
(\bibinfo{year}{2018}).

\bibitem[{\citenamefont{{S. Latini}}(2017)}]{wd21}
\bibinfo{author}{\bibnamefont{{S. Latini, Kirsten T. Winther, T. Olsen, and K. S. Thygesen}}}, \bibinfo{journal}{Nano Lett.} \textbf{\bibinfo{volume}{17}}, \bibinfo{pages}{938}
(\bibinfo{year}{2017}).

\bibitem[{\citenamefont{{Kristen Kaasbjerg}}(2012)}]{wd22}
\bibinfo{author}{\bibnamefont{{K. Kaasbjerg, K. S. Thygesen, and K. W. Jacobsen}}}, \bibinfo{journal}{Phys. Rev. B} \textbf{\bibinfo{volume}{85}} \bibinfo{pages}{115317} (\bibinfo{year}{2012}).

\bibitem[{\citenamefont{{K. Kaasbjerg}}(2014)}]{wd23}
\bibinfo{author}{\bibnamefont{{K. Kaasbjerg, K. S. Bhargavi, and S. S. Kubakaddi}}},  \bibinfo{journal}{Phys. Rev. B} \textbf{\bibinfo{volume}{90}} \bibinfo{pages}{165436} (\bibinfo{year}{2014}).

\bibitem[{\citenamefont{{A. Thilagam}}(2016)}]{wd24}
\bibinfo{author}{\bibnamefont{{A. Thilagam}}}, \bibinfo{journal}{ J. Appl. Phys.} \textbf{\bibinfo{volume}{120}} \bibinfo{pages}{124306} (\bibinfo{year}{2016}).

\bibitem[{\citenamefont{{T. D. Lee}}(1953)}]{wd25}
\bibinfo{author}{\bibnamefont{{T. D. Lee, F. E. Low, and D. Pines}}}, \bibinfo{journal}{Phys. Rev.} \textbf{\bibinfo{volume}{90}}, \bibinfo{pages}{297}
(\bibinfo{year}{1953}).

\bibitem[{\citenamefont{{J. Pollmann}}(1977)}]{wd26}
\bibinfo{author}{\bibnamefont{{J. Pollmann and H. B$\ddot{u}$ttner}}}, \bibinfo{journal}{Phys. Rev. B} \textbf{\bibinfo{volume}{16}}, \bibinfo{pages}{4480}
(\bibinfo{year}{1977}).

\bibitem[{\citenamefont{{I. Kyl}}(2015)}]{wd27}
\bibinfo{author}{\bibnamefont{{I. Kyl$\ddot{a}$np$\ddot{a}$$\ddot{a}$ and H. P. Komsa}}}, \bibinfo{journal}{Phys. Rev. B} \textbf{\bibinfo{volume}{92}}, \bibinfo{pages}{205418}
(\bibinfo{year}{2015}).

\bibitem[{\citenamefont{{Z. Jin}}(2014)}]{wd28}
\bibinfo{author}{\bibnamefont{{Z. Jin, X. Li, J. T. Mullen, and K. W. Kim}}}, \bibinfo{journal}{Phys. Rev. B} \textbf{\bibinfo{volume}{90}}, \bibinfo{pages}{045422}
(\bibinfo{year}{2014}).

\bibitem[{\citenamefont{{S. Ovesen}}(2018)}]{wd29}
\bibinfo{author}{\bibnamefont{{S. Ovesen, S. Brem, C. Linder, M.
Kuisma, P. Erhart, M. Selig, and E. Malic}}}, \bibinfo{journal}{arXiv:1804.08412v1}
(\bibinfo{year}{2018}).

\bibitem[{\citenamefont{{S. Tongay}}(2012)}]{wd30}
\bibinfo{author}{\bibnamefont{{S. Tongay, J. Zhou, C. Ataca, K. Lo, T. S. Matthews, J. Li, J. C. Grossman and J. Wu}}}, \bibinfo{journal}{Nano. Lett.} \textbf{\bibinfo{volume}{12}}, \bibinfo{pages}{5576} (\bibinfo{year}{2012}).

\bibitem[{\citenamefont{{Jason. S. Ross}}(2013)}]{wd31}
\bibinfo{author}{\bibnamefont{{Jason. S. Ross, S. F. W, H. Y. Yu, N. J. Ghimire, A. M. Jones, G. Aivazian, J. Q. Yan, D. G. Mandrus, D. Xiao, W. Yao and X. Xu}}}, \bibinfo{journal}{Nat. Commun.} \textbf{\bibinfo{volume}{4}}, \bibinfo{pages}{1474} (\bibinfo{year}{2013}).

\bibitem[{\citenamefont{{T. Sohier}}(2016)}]{wd32}
\bibinfo{author}{\bibnamefont{{T. Sohier, M. Calandra, and F. Mauri}}}, \bibinfo{journal}{Phys. Rev. B} \textbf{\bibinfo{volume}{94}}, \bibinfo{pages}{085415}
(\bibinfo{year}{2016}).

\end{thebibliography}
\end{document}